\documentclass[aps,amssymb,amsmath,singlecolumn, pra]{revtex4}
\usepackage{graphicx}
\usepackage{epsfig}
\usepackage{dcolumn}% Align table columns on decimal point
\usepackage{bm}% bold math
\usepackage{bbm}
\usepackage[up]{subfigure}
\newcommand{\be}{\begin{equation}}
\newcommand{\ee}{\end{equation}}
\newcommand{\bc}{\begin{center}}
\newcommand{\ec}{\end{center}}
\newcommand{\bea}{\begin{eqnarray}}
\newcommand{\eea}{\end{eqnarray}}

\begin{document}
% Title & Authors
\title{Discrete time quantum walk model for single and entangled particles to retain entanglement in coin space}
%\date{\today}
%\date[]{October 10, 2006}
%\date[\bf DRAFT:\rm\ ]{-13-Sep-06}
\author{C. M. \surname{Chandrashekar}}
\affiliation{Institute for Quantum Computing, University of Waterloo,\\
200 University Avenue West, Waterloo, Ontario N2L 3G5, Canada}
%==================================================

\begin{abstract}

%==================================================

In most widely discussed discrete time quantum walk model, after every unitary shift operator, the particle evolves into the superposition of position space and settles down in one of its basis states, loosing entanglement in the coin space in the new position. The Hadamard operation is applied to let the particle to evolve into the superposition in the coin space and the walk is iterated. We present a model with a additional degree of freedom for the unitary 
shift operator $U^{\prime}$. The unitary operator with additional degree of freedom 
will evolve the quantum particle into superposition of position space retaining the entanglement in coin space. This eliminates the need for quantum coin toss (Hadamard operation) after every unitary displacement operation as used in most widely studied version of the discrete time quantum walk model. This construction is easily extended to a multiple particle quantum walk and in this article we extend it for a pair of particles in pure state entangled in coin degree of freedom by simultaneously subjecting it to a pair of unitary displacement operators which were constructed for single particle.  We point out that unlike for single particle quantum walk, upon measurement of its position after $N$ steps, the entangled particles are found together with $1/2$ probability and at different positions with $1/2$ probability. This can act as an advantage in applications of the quantum walk. A special case is also treated using a complex physical system such as, inter species two-particle entangled Bose-Einstein condensate, as an example.
\end{abstract}

%====================================================
\maketitle
\preprint{Version}
%====================================================
\section{Introduction}
\label{intro}
%====================================================

Quantum walks, the counterpart of classical random walks for particles which cannot be precisely localized due to quantum uncertainties was first introduced in
1993~\cite{aharonov}. As classical random walk has played a very prominent role in the classical information theory, quantum walk has been widely explored for its role in the quantum information theory. Several quantum algorithms with optimal efficiency has been proposed~\cite{childs, shenvi, childs1, ambainis}. Various schemes have been proposed for the physical realization of quantum walks~\cite{travaglione, rauss, eckert, chandra1}. Experimental implementations have also been reported~\cite{ryan,du}.

A discrete time and a continuous time models of quantum walk have been widely studied
using a single quantum walker~\cite{kempe}. In the most widely referred literature, the discrete time quantum walk have been constructed using a coin toss instruction and while the continuous time do not require coin toss. In Ref.~\cite{patel} discrete time quantum walk without a coin toss instruction has been defined and analyzed. The evolution operators (Laplacian) that are unitary and ultra local has been discretized. The Laplacian operator has been decomposed into even and odd parts. This construction of discrete time quantum walk has exchanged the coin states for the even-odd site labels. However the above model reports the loss of short distance homogeneity, i.e., the translational invariance holds in steps of 2 instead of steps of 1. 

In most widely discussed discrete time quantum walk model, after every unitary shift operator, the the particle evolves into the superposition of position space and settles down in one of its basis states, loosing entanglement in the coin space in the new position. The Hadamard operation is applied to let the particle to evolve into the superposition in the coin space. This process
of unitary operation followed by Hadamard operation is iterated to realize large number of steps of the quantum walk. In our model we construct the unitary shift operator, $U^{\prime}$, by adding an additional degree of freedom.  
The Application of the unitary shift operator $U^{\prime}$ on the particle entangled in coin space, $\frac{1}{\sqrt{2}} (|0, x_{0}\rangle + i |1,x_{0}\rangle)$, where 0 and 1 are the two states of the particle at position, $x_{0}$, before the beginning of the quantum walk is retained after every unitary shift operation. This eliminates the need for quantum coin toss (Hadamard operation) after every unitary displacement operation as used in most widely studies version of the discrete time quantum walk. 

Entanglement is one of the important properties of quantum mechanics. It lies in the heart of the profound difference between quantum mechanics and classical physics~\cite{bell-book}. Recently, much attention has been paid to the applications of entanglement in the emerging field of quantum information processing. Role of entanglement in quantum walks has also been explored. In Ref.~\cite{omar}, a two particle discrete time quantum walker has been discussed, in which two non-interacting, distinguishable particle in a pure separable state and two maximally entangled state has been discussed separately using the quantum coin toss. In Ref.~\cite{andraca}, mathematical formalism for un-restricted quantum walks with entangled coins and one walker has been presented. Keeping the role of entanglement into consideration we extent the construction of unitary operator to pair of two-state entangled particles.

We point out that the discrete time quantum walk in a line using a pair of entangled particles, when simultaneously subjected to a pair of unitary shift operators $U^{\prime}$ constructed for single particle, shifts the state of the entangled particles in superposition of position space retaining the entanglement in the coin space. This leaves the entangled particle quantum walk to evolve without a need for the quantum coin toss instruction (Hadamard operation) after every unitary operator as in widely studied version of discrete time quantum walk.
We also present some interesting features of quantum walk using two two-state entangled particle. In our construction, we initially consider a pair of particle in pure state entangled in coin degree of freedom at position $x_{0}$, $\frac{1}{\sqrt{2}} (|0, x_{0}\rangle_{1} |1, x_{0}\rangle_{2} + i |1,x_{0}\rangle_{1} |0, x_{0}\rangle_{2})$, where the subscripts indicate the label of the first and the second particle. The construction can be generalized for all entangled states of two two-state particles. We point out that upon measurement after implementing $N$ number of steps of the quantum walk, the entangled particles are found together with $1/2$ probability and at different positions with $1/2$ probability. This chance of registering the individual particles separately will reduce the number of runs required to experimentally confirm the quantum behavior of the walk by $1/3$. This can act as an advantage in applications of the quantum walk. We also discuss the way unitary operator can be modified for the quantum walk using a specific systems, such as Bose-Einstein condensate which occurs in an inter species two-particle entangled state. 

Lets start with the simple comparison of the quantum walk with the classical random walk and then introduce to the version of the discrete time quantum walk, studied in the literature.

%==========================================
\section{Discrete time quantum walk}
\label{discrete}
%==========================================

In one-dimensional classical random walk, to make a step of a given length to the left or to the right for a particle are described in terms of probabilities. On the other hand, in quantum walk they are described in terms of probability amplitudes. In an unbiased one-dimensional classical random walk, with the particle initially at $x_{0}$, evolves in such a way that at each step, the particle moves with probability $1/2$ one step to the left and with probability $1/2$ one step to the right. In a quantum mechanical analog the state of the particle evolves at each step into a coherent superposition of moving one step to the right and one step to the left with equal probability amplitude.

To realize the commonly studied version of discrete time quantum (Hadamard) walk, two
degrees of freedom, the internal state, which is called as {\it coin} Hilbert space
$\mathcal H_{c}$ (quantum coin)  and the {\it position} Hilbert space $\mathcal H_{p}$
is required. Imagine $\mathcal H_{c}$ of a particle on a line spanned by two basis
states $|0\rangle$ and $|1\rangle$, and $\mathcal H_{p}$ as spanned by basis states
$|x\rangle$:$x$={\bf Z}. The state of the total system is in the space $\mathcal H=
\mathcal H_{c} \otimes \mathcal H_{p}$. The internal state of the particle
determines the direction of the particles movement when the conditional unitary shift
operator $U$ is applied.

Conditioned unitary shift operator $U$ on the particle with internal state being
$|0\rangle$ ($|1\rangle$), the particle moves to the left (right), i.e.,
$U(|0\rangle\otimes|x\rangle) =|0\rangle\otimes|x-1\rangle$ and
$U(|1\rangle\otimes|x\rangle) =|1\rangle\otimes|x+1\rangle$.

The particle initially in state $|0\rangle$ ($|1\rangle$) at $x_{0}$ is made to evolve into the superposition state by applying the Hadamard rotation $H$, 
\begin{equation} 
H =\frac{1}{\sqrt 2} \left( \begin{array}{clcr}
 1  & &   1   \\
 1  & &  -1
 \end{array} \right), \end{equation} \noindent such that, \begin{eqnarray}
\label{hadamard} (H\otimes\mathbbm{1})|0,x\rangle &=& \frac{1}{\sqrt 2}[|0,x\rangle+|1,
x\rangle] \nonumber \\ (H\otimes \mathbbm{1})|1, x\rangle &=& \frac{1}{\sqrt 2}[|0,
x\rangle-|1, x\rangle]. \end{eqnarray} A unitary shift operator $U$, \be
\label{eq:condshift} U =|0\rangle \langle 0|\otimes \sum |x-1\rangle \langle x
|+|1\rangle \langle 1 |\otimes \sum |x+1\rangle \langle x|, \ee on the particle whose
wave function is given by $|\Psi_{in}\rangle=\frac{1} {\sqrt 2}[|0\rangle \pm
|1\rangle]\otimes|\Psi_{x_{0}}\rangle$, is applied to move the particle in superposition
of position space. $U$ can also be written as, \be \label{eq:alter} U =
\exp(-2iS\otimes Pl), \ee \noindent $P$, being the momentum operator and $S$,
the operator corresponding to the state of the particle respectively. The
eigenstates of $S$ are denoted by $|0\rangle$ and $|1\rangle$ and $l$, corresponds 
to one step length. 

Therefore, \begin{equation}
\label{eq:uprim} U|\Psi_{in}\rangle=\frac{1}{\sqrt 2}[|0\rangle\otimes e^{-iPl}
\pm |1\rangle\otimes e^{iPl}]|\Psi_{x_{0}}\rangle. \end{equation}

Application of $U$ on $|\Psi_{in}\rangle$ spatially entangles the $\mathcal H_{c}$ and
$\mathcal H_{p}$. After every unitary shift operator the particle settles in one of its
internal state and hence the particle is again made to evolve into the superposition
state by applying the Hadamard rotation $H$. Each step of quantum (Hadamard) walk is
composed of unitary shift operation followed by the Hadamard operation (rotation) $H$. The process is iterated without resorting to the intermediate measurement to realize large number of steps.

The probability amplitude distribution arising from the iterated application of
$W=U(H\otimes \mathbbm{1})$ is significantly different from the distribution of the
classical walk after the first two steps~\cite{kempe}.  If the coin initially is in a
suitable superposition of $|0\rangle$ and $|1\rangle$ then the probability amplitude
distribution after $n$ steps of quantum walk will have two maxima symmetrically
displaced from the starting point. The variance of quantum version grows quadratically
with number of steps $n$, $\sigma^{2}\propto n^{2}$ compared to $\sigma^{2}\propto n$
for the classical random walk.

%==================================
\section{Unitary shift operator to retain entanglement in coin space}
\label{mod-unitary}
%===============================

The Unitary shift operator $U$ defined in Eq. (\ref{eq:alter}) can also be written as ~\cite{kempe}, 
\be 
\label{eq:alter1} 
U = e^{-2iS\otimes Pl} = e^{-i(|0\rangle \langle 0| - |1\rangle \langle 1|)\otimes Pl} =(|0\rangle \langle 0|\otimes e^{-iPl})(|1\rangle \langle 1|\otimes e^{iPl}),
\ee 
where $P$ being the momentum operator and $S$, the operator corresponding to the state of the particle respectively. 
To implement the discrete time quantum walk without loosing the superposition in coin space after every unitary operator, along with the two degree of freedom,
{\it coin} Hilbert space $\mathcal H_{c}$ and the {\it position} Hilbert space $\mathcal H_{p}$, we introduce a third  degree of freedom {\it momentum} Hilbert space $\mathcal H_{m}$. The momentum Hilbert space, $\mathcal H_{m}$, of the momentum operator is spanned by two basis states $|0\rangle_{o}$ and $|1\rangle_{o}$.    

The state of the total system now is in the space $\mathcal H = \mathcal H_{c} \otimes (\mathcal H_{m} \otimes \mathcal H_{p}) $. The internal state of the particle and the state of the momentum operator determines the direction of the particle movement when the conditional unitary shift operator $U^{\prime}$ is applied.

Momentum operator $P$ is denoted by the eigenstates $|0\rangle_{o}$ and $|1\rangle_{o}$. That is, momentum operator exists in  the coherent superposition state. The momentum operator is defined such that, if it is in state $|0\rangle_{o}$, it will shift the particle in state $|0\rangle$ to the left and particle in state $|1\rangle$ to the right. Similarly if the momentum operator is in state $|1\rangle_{o}$, it will shift the particle in state $|0\rangle$ to the right and particle in state $|1\rangle$ to the left. Therefore the modified version of Eq. (\ref{eq:alter1}) can be written as, 
\be 
\label{eq:alter2} 
U^{\prime} = (|0\rangle \langle 0|\otimes e^{-i(|0\rangle_{o} \langle 0|_{o} - |1\rangle_{o} \langle 1|_{o})\otimes pl})(|1\rangle \langle 1|\otimes e^{i(|0\rangle_{o} \langle 0|_{o} - |1\rangle_{o} \langle 1|_{o})\otimes pl})
\ee 
\be
\label{eq:alter3}
U^{\prime} = [|0\rangle \langle 0| \otimes \{(|0\rangle_{o} \langle 0|_{o}\otimes e^{-ipl})(|1\rangle_{o} \langle 1|_{o}\otimes e^{ipl})\}][|1\rangle \langle 1| \otimes \{(|0\rangle_{o} \langle 0|_{o}\otimes e^{ipl})(|1\rangle_{o} \langle 1|_{o}\otimes e^{-ipl})\}].
\ee 
This can be written as,
\be 
\label{eq:alter4} 
U^{\prime} = (|0\rangle \langle 0|\otimes U_{a})(|1\rangle \langle 1|\otimes U_{b})
\ee 
where,
\be
\label{ua}
 U_{a}= (|0\rangle_{o} \langle 0|_{o}\otimes e^{-ipl})(|1\rangle_{o} \langle 1|_{o}\otimes e^{ipl})
\ee 
\be
\label{ub}
U_{b} = (|0\rangle_{o} \langle 0|_{o}\otimes e^{ipl})(|1\rangle_{o} \langle 1|_{o}\otimes e^{-ipl}).
\ee
If the initial wave function of the particle and momentum operator are given by $|\Psi_{in}\rangle=\frac{1} {\sqrt 2}[|0\rangle \pm |1\rangle]\otimes|\Psi_{x_{0}}\rangle$ and $|\Phi_{in}\rangle=\frac{1} {\sqrt 2}[|0\rangle_{o} \pm |1\rangle_{o}]\otimes|\Phi_{x_{0}}\rangle$ respectively, then the operation of $U^{\prime}$ on this system can be written as,

\begin{equation}
\label{eq:uprim2} U^{\prime}(|\Psi_{in}\rangle \otimes |\Phi_{in}\rangle) =\frac{1}{\sqrt 2}[|0\rangle\otimes 
\frac{1}{\sqrt 2}(|0\rangle_{o} \otimes e^{-ipl} \pm |1\rangle_{o}\otimes e^{ipl})
\pm |1\rangle \otimes \frac{1}{\sqrt 2} (|0\rangle_{o} \otimes e^{ipl} \pm |1\rangle_{o}\otimes e^{-ipl})](|\Psi_{x_{0}}\rangle \otimes |\Phi_{in}\rangle). 
\end{equation}

\begin{equation}
\label{eq:uprim3} U^{\prime}(|\Psi_{in}\rangle \otimes |\Phi_{in}\rangle) =\frac{1}{\sqrt 4}[|0\rangle\otimes 
|0\rangle_{o} \otimes e^{-ipl} \pm |0\rangle \otimes |1\rangle_{o}\otimes e^{ipl}
\pm |1\rangle \otimes |0\rangle_{o} \otimes e^{ipl} \pm |1 \rangle \otimes |1\rangle_{o}\otimes e^{-ipl})](|\Psi_{x_{0}}\rangle \otimes |\Phi_{in}\rangle). 
\end{equation}

The third degree of freedom, momentum Hilbert space can be identified with the position space and therefore, $U^{\prime}$ can be simplified and written with only particle state and the position space components as,

\begin{eqnarray}
\label{new-unitary}
U^{\prime} = \{|0\rangle \langle 0|\otimes  \frac {1}{\sqrt 2}(\sum |x-1\rangle \langle x | \pm \sum |x+1\rangle \langle x|) + |1\rangle \langle 1|\otimes  
\frac{1}{\sqrt 2} (\sum |x+1\rangle \langle x | \pm \sum |x-1\rangle \langle x|).
\end{eqnarray}

The state of the particle after the first step of the quantum walk on application of the unitary operator $U^{\prime}$ can be written as,
\begin{eqnarray}
\label{new-unitary1}
U^{\prime}|\Psi_{x_{1}}\rangle =  \frac {1}{\sqrt 4} [(|0\rangle \pm |1\rangle) \otimes |x-1\rangle \pm (|0\rangle \pm |1\rangle) \otimes |x+1\rangle]|\Psi_{x_{0}}\rangle.
\end{eqnarray}

From the Eq. (~\ref{new-unitary1}) one can conclude that the unitary operator $U^{\prime}$ on particle entangled in the coin space evolves into the superposition of the position space retaining the entanglement in the coin space. This will eliminate the need for Hadamard operation after every unitary shift operation as in most widely studied version of the discrete time quantum walk. The $N$ step of the quantum walk can be implemented by applying $U^{\prime N}$ on the particle initially in the  superposition of the coin space.

The state of the momentum operator is treated to be in the coherent superposition of its eigen states. In this construction, the walk can be biased by having different probability amplitudes of the states of the momentum operator along with states of the particle giving more freedom to manipulate the quantum walk (bias the walker).

One can consider a simple physical system such as polarized light as momentum operator. It can be conditioned such that, the vertically polarized light will 
shift the particle in state $|0\rangle$ to the left and particle in state $|1\rangle$ to the right. The  horizontally polarized light shift the particle in state $|1\rangle$ to the left and particle in state $|0\rangle$ to the right. Considering the light in coherent superposition of the vertical and horizontal polarization as the momentum operator, one can implement unitary operator $U^{\prime}$ as discussed in this section.  

%================================================
\section{Quantum Walk with a pair of entangled particles}
\label{entangled-particle}
%================================================
Let us extend the quantum walk model constructed in the previous section to the two entangled particle pair on a line, two two-state particles entangled in coin degree of freedom in this section. The two particles need not have to be identical. The Hilbert space of our paired entangled particle system is given by: 
\be 
\label{hilbert} 
\mathcal H_{E} \equiv  \mathcal H_{1} \otimes \mathcal H_{2} \equiv  \{\mathcal H_{c1} \otimes (\mathcal H_{m1} \otimes \mathcal H_{p1})\} \otimes  \{\mathcal H_{c2} \otimes (\mathcal H_{m2} \otimes \mathcal H_{p2})\},
\ee 
where $\mathcal H_{1}$ and $\mathcal H_{2}$ represent the Hilbert spaces for particles 1 and 2, respectively. We find a basis of four orthogonal, maximally entangled states, the so called Bell-states basis to be, 
\be \label{bell1} |\psi^{\pm}_{0}\rangle_{12} = \frac{1}{\sqrt 2}[|0\rangle_{1}|1\rangle_{2} \pm |1\rangle_{1} |0\rangle_{2}], 
\ee 
\be 
\label{bell2}
|\Phi^{\pm}_{0}\rangle_{12} = \frac{1}{\sqrt 2}[|0\rangle_{1}|0\rangle_{2} \pm
|1\rangle_{1} |1\rangle_{2}]. 
\ee 
Lets consider a case where both particles start the quantum walk in the same position but with different coin state. We will consider a pair of entangled particles with complex amplitude at position $x_{0}$,
\be 
\label{ent-pos0} 
|\psi_{0}\rangle_{12} = \frac{1}{\sqrt 2}[|0, x_{0}\rangle_{1}|1, x_{0}\rangle_{2} + i |1, x_{0}\rangle_{1}|0, x_{0}\rangle_{2}].
\ee 
The complex amplitude is chosen in the initial state only to keep the real and imaginary part separate during the analysis. Each step of the quantum walk on this system of entangled pair of particles can be implemented by simultaneously applying a pair of unitary shift operator $U^{\prime}$. The complete unitary shift operator on the system is of the form,
\be
U^{\prime}_{E} = U^{\prime} \otimes U^{\prime},
\ee
where $U^{\prime}$ is given by Eq. (\ref{new-unitary}) and is same for both the particles. 
Therefore the unitary shift operator on the entangled particle pair can be written as,
\begin{eqnarray} 
\label{unitary-evolu}
U^{\prime}_{E}|\psi_{x}\rangle_{12} = \frac{1}{\sqrt 2} (U^{\prime}|0, x\rangle_{1}
U^{\prime}|1, x\rangle_{2} + i U^{\prime} |1 , x\rangle_{1} U^{\prime} |0 , x\rangle_{2}).
\end{eqnarray}
\begin{eqnarray} 
\label{unitary-evolu1}
U^{\prime}|\psi_{x}\rangle_{12} =\frac{1}{\sqrt 8}[(|0,x-1\rangle_{1} |1,x-1\rangle_{2}+ |0,x-1\rangle_{1}|1,x+1\rangle_{2} + |0,x+1\rangle_{1} |1,x+1\rangle_{2} + 
|0,x+1\rangle_{1} |1,x-1\rangle_{2}) + \nonumber \\
i (|1,x+1\rangle_{1} |0,x+1\rangle_{2} +  |1,x+1\rangle_{1} |0,x-1\rangle_{2} + 
|1,x-1\rangle_{1} |0,x-1\rangle_{2} + |1,x-1\rangle_{1} |0,x+1\rangle_{2})].
\end{eqnarray}
The expression can be re-organized and written as,
\begin{eqnarray} 
\label{end-posx1} 
|\psi_{x_{1}}\rangle_{12} = \frac{1}{\sqrt 8}[(|0, x-1\rangle_{1}|1, x-1\rangle_{2} + i |1, x-1\rangle_{1} |0, x-1\rangle_{2}) + 
(|0, x+1\rangle_{1}|1, x+1\rangle_{2} + i |1, x+1\rangle_{1} |0, x+1\rangle_{2})+ \nonumber \\
(|0, x-1\rangle_{1}|1, x+1\rangle_{2} + i |1, x-1\rangle_{1} |0, x+1\rangle_{2}) + 
(|0, x+1\rangle_{1}|1, x-1\rangle_{2} + i |1, x+1\rangle_{1} |0, x-1\rangle_{2})].\end{eqnarray} 

The above operation will evolve the initial states to the new position space $(x+1)$ 
and $(x-1)$ retaining the entanglement in coin degree of freedom of the two particles. The final state of the system after $N$ steps will be:
\begin{eqnarray} 
\label{nsteps} 
U^{\prime N}_{E}|\psi_{0}\rangle_{12} = 
= \frac{1}{\sqrt 2} (U^{\prime N}|0, x\rangle_{1}
U^{\prime N}|1, x\rangle_{2} + i U^{\prime N} |1 , x\rangle_{1} U^{\prime N} |0 , x\rangle_{2}).
\end{eqnarray}

\begin{figure}
\begin{center}
\epsfig{figure=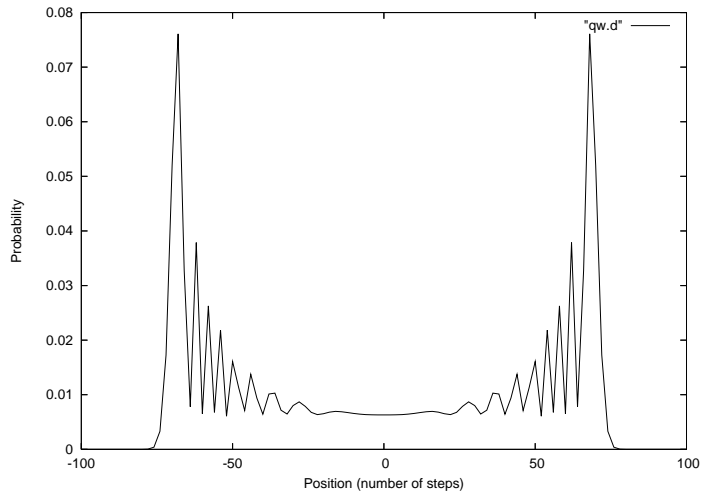, height= 9.4cm, angle=-90}
\caption{\label{qrw-ent100}  Probability distribution of
the quantum walk using pair of entangled particles starting with
the initial state $\frac{1}{\sqrt 2}[|0, x_{0}\rangle_{1}|1, x_{0}\rangle_{2} + i|1, x_{0}\rangle_{1} |0, x_{0}\rangle_{2}]$. The distribution is for 100 steps.}
\end{center}
\end{figure}

Therefore, the conditioned unitary shift operator $U^{\prime}_{E}$, the pair of unitary operator $U^{\prime}$ applied simultaneously, entangles the coin space $\mathcal H_{c}$ (quantum coin) with the position Hilbert space $\mathcal H_{p}$ retaining the entanglement in coin space after each unitary operation. 
From the above description, one can conclude that the pair of entangled particle in state of the form, Eq. (\ref{ent-pos0}), when simultaneously subjected to pair of unitary operators  $U^{\prime}$ behaves as a quantum walk described in literature. From the Fig. (\ref{qrw-ent100}) one can notice the symmetric probability distribution of finding the entangled particles after it has undergone a quantum walk of 100 steps. 

One can easily verify that the unitary operator $U^{\prime}_{E}$ can also be used in the same way as described above to implement the quantum walk on the maximally entangled states, Bell states, of the form, Eq. (\ref{bell1}) and Eq. (\ref{bell2}).

%=====================================
\section{Entangled particle probability distribution} 
\label{ent}
%==========================================

If one happens to measure the position of the two particles immediately after the first step, it can be noted from the Eq. (\ref {end-posx1}) that, the particle $1$ and the particle $2$ can be seen together at $(x-1)$ or at $(x+1)$ with probability $1/2$ and can be seen separately at $(x-1)$ and $(x+1)$ with probability $1/2$. 
That is, the quantum walk with pair of entangled particles using unitary operator of the form $U^{\prime}_{E}$, upon measurement after implementing quantum walk of $N$ steps, one can find the particle $1$ and particle $2$ at the same position or at different positions with equal probability. 
\be
P_{N}(1_{i}, 2_{j}) = 0.5
\ee
\be
P_{N}(1_{i}, 2_{i}) = 0.5
\ee
where, $P_{N}(1_{i}, 2_{j})$ is the probability of the particle $1$ and the particle $2$ being in two different position $i$ and $j$ after $N$ steps and $P_{N}(1_{i}, 2_{i})$ is the probability of particle $1$ and particle $2$ being at the same position $i$.  

To obtain the probability distribution of the position of the single particle 
quantum walker after $N$ steps experimentally, the position of the walker is 
registered after $N$ steps and the process is iterated for  $n >> N$ times. 
The $n$ position points is then used to obtain the distribution. While for the 
quantum walk using two entangled particles, upon measurement after $N$ steps, 
with $1/2$ probability the particle $1$ and the particle $2$ are registered at 
the same position and with an other $1/2$ probability the two particles are 
registered at two different positions adding on to the data of position of the 
particles. Therefore after iterating the $N$ steps of the quantum walk just 
$\frac{2}{3}n$ times one can have $n$ position points registered. This reduces 
the number of runs necessary to conclude the quantum behavior of the two 
two-state entangled particle walker. This can be of an advantage for the applications of the quantum walk. 

Along with preserving the superposition in the coin space, the number of iterations of $N$ steps of the quantum walk necessary to conclude the quantum behavior of the pair of entangled particle walk is $1/3$ times lesser than the number of runs required by other earlier proposals. By increasing the number of entangled particles then number of iteration required can be further reduced. This also makes the practical implementation easier and be an advantage for applications of the quantum walk.

%================================================
\section{Implementation in an Entangled physical system}
\label{special-case}
%================================================

An entangled paired photons, entangled paired fermions or bosons can be a suitable physical system where the quantum walk model presented for entangled particles in Sec. \ref{entangled-particle} can be tested. A inter species two-particle entangled Bose-Einstein condensate~\cite{shi}, may be used to realize the construction presented in this article in macroscopic scale. Depending on the physical system, the physical implementation of the of unitary operator appropriate for the system can be worked out. A scheme for the implementation of quantum walk using Bose-Einstein condensate (BEC) has been suggested in ~\cite{chandra1}. In this section we will develop on that and outline a scheme to implement the quantum walk without coin toss instruction in regular interval for a complex system such as, inter species two-particle entangled Bose-Einstein condensate (BEC)~\cite{shi}.  

If the two species entangled particles in the BEC are spatially separated, the particles will evolve out of the coherent state, and hence will no more be identified as a BEC. If sufficient measures are taken to retain the coherence (BEC state) in the two-particle entangled BEC upon application of the unitary operator, the implementation of the quantum walk without need for the Hadamard operation can be constructed. 

In Ref.~\cite{chandra1}, a stimulated Raman process are used to drive transition between two optically trappable states $|0\rangle$ and $|1\rangle$  using the virtual state $|e\rangle$ as an intermediate state and impart a well defined momentum to spatially translate the BEC in Schr\"odinger cat state. A unitary shift operation, $U$, called as a {\it stimulated Raman kicks} can be applied on the BEC using stimulated Raman process. A pair of counter-propagating laser beams 1 and 2  with frequency $\omega_{1}$ and $\omega_{2}$ and a wave-vector $k_{1}$ and $k_{2}$ is applied on the BEC for a 2N-photon transition time (N is the number of atoms in the BEC) to implement one unitary shift operation. These beams are configured to propagate along the axial direction of the optical dipole trap. A stimulated Raman transition occurs when an atom changes its state by coherently exchanging photons between the two laser fields, absorption of photon from laser field 1 and stimulated emission into laser field 2 or by absorption from field 2 and stimulated emission into field 1.  The BEC initially in eigen state $|0\rangle$ ($|1\rangle$) can absorb photon from field 1 (2) and re-emit photon into field 2 (1). This inelastic stimulated Raman scattering process imparts well-defined momentum, stimulated Raman kick, on the BEC.

For the inter species two-particle entangled BEC, one needs a minimum of two pair of counterpropagating beams (one pair for each species) to implement a unitary operation on the system. But two pair of counterpropagating laser beams are not sufficient to eliminate the need for the Hadamard operation after every unitary operation.  

A four pair of counterpropagating laser beams, two identical laser beams each of frequency $\omega_{1}$, $\omega_{2}$, $\omega_{3}$ and $\omega_{4}$ are configured to counterpropagate each other. Laser beams of frequency $\omega_{1}$ and $\omega_{2}$ are the transition frequency for particle $1$ and $\omega_{3}$ and $\omega_{4}$ are the transition frequency for the particle $2$. Unlike in Ref.~\cite{chandra1}, each beam are configured to counterpropagate. If the particle $1$ absorbs frequency $\omega_{1}$ and emit frequency (stimulated emission) $\omega_{2}$ the particle gets a kick to the left or to the right depending on direction of the laser beam of frequency $\omega_{1}$ it initially absorbs, as it propagate in both the directions. This along with moving the particle to the left and right it can also absorb and re-emit light of the same frequency ($\omega_{1}$) and hence giving no physical displacement. It would be same for absorption of frequency $\omega_{2}$. In the same way $\omega_{3}$ and $\omega_{4}$ are configured for transition frequency of the particle $2$.

Other possible combinations of absorption and stimulated emission are supressed when the bonding between the entangled particle are stronger and act as a constrain for the particles $1$ and $2$ to be physically separated. And they have to 
remain together to retain the BEC state. 

We can define $U_{1}$ and $U_{2}$ as the unitary operators for the particle $1$ and the particle $2$ respectively. In Ref. \cite{chandra1} it has been pointed out that during the stimulated Raman kicks to implement unitary operator the internal state of the particle changes. In Ref. \cite{chandra2} it has been pointed out that the quantum walk remains invariant, except for a spatial inversion if the unitary shift operator induce a bit flip on the state of the particle. For the laser beam configurations as suggested in this section one can write down the unitary operators as below,
\begin{eqnarray} \label{unitary-bec}
U_{1} (|0, x\rangle)_{1} =\frac{1}{\sqrt 4}[|1, x-1\rangle_{1} + |1, x+1\rangle_{1} + 2 |0, x\rangle_{1}]\nonumber \\
U_{1}(|1, x\rangle)_{1} =\frac{1}{\sqrt 4}[|0, x-1\rangle_{1} + |0, x+1\rangle_{1} + 2|1, x\rangle_{1}]\nonumber \\
U_{2} (|0, x\rangle)_{2} =\frac{1}{\sqrt 4}[|1, x-1\rangle_{2} + |1, x+1\rangle_{2} + 2 |0, x\rangle_{2}]\nonumber \\ 
U_{2}(|1, x\rangle)_{2} =\frac{1}{\sqrt 4}[|0, x-1\rangle_{2} + |0, x+1\rangle_{2} + 2|1, x\rangle_{2}].
\end{eqnarray}
\begin{figure}
\begin{center}
\epsfig{figure=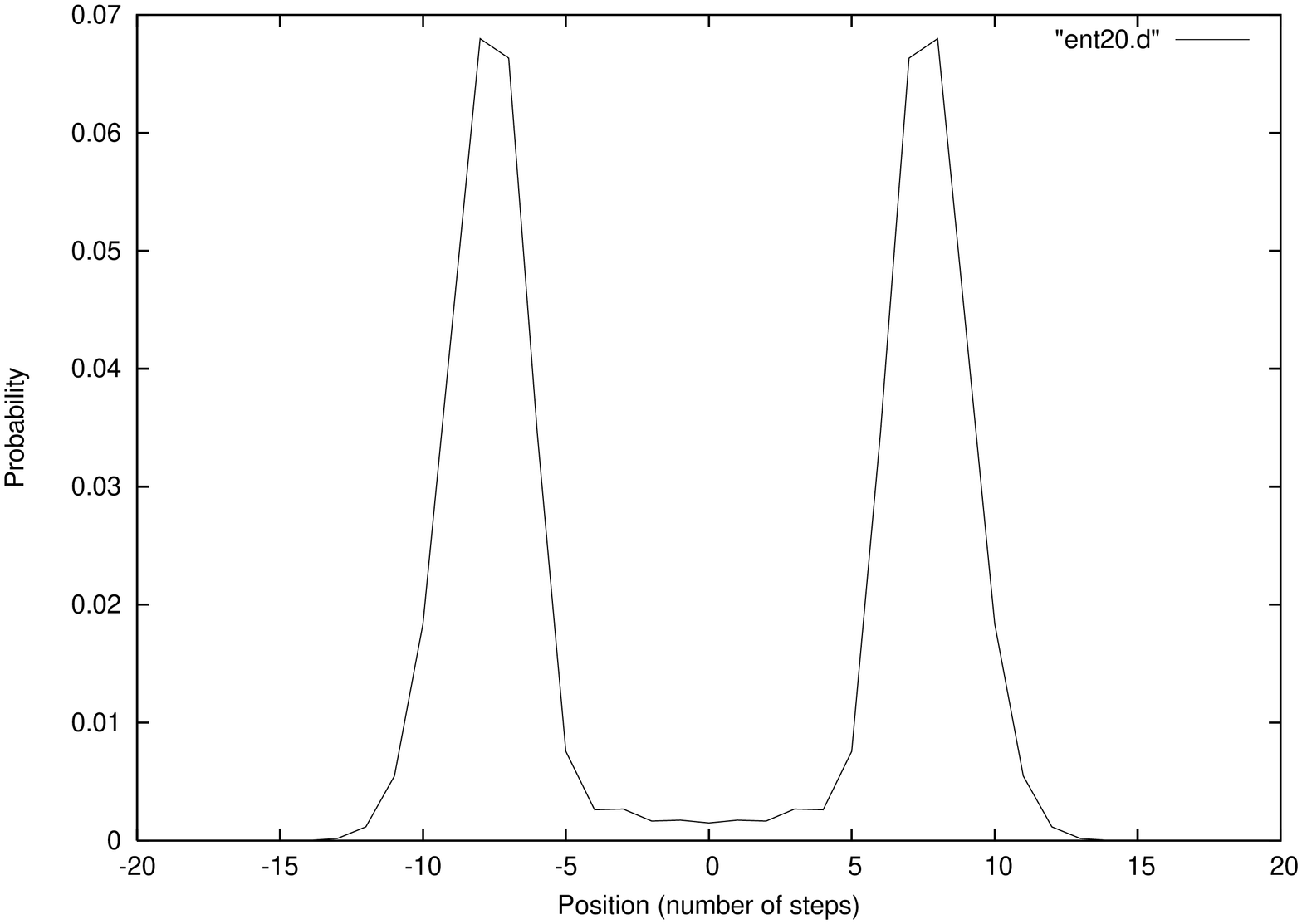, height= 9.4cm, angle=-90}
\caption{\label{entbec20}  Probability distribution of
the constrained quantum walk using entangled particle pair. The distribution is for 20 steps.}
\epsfig{figure=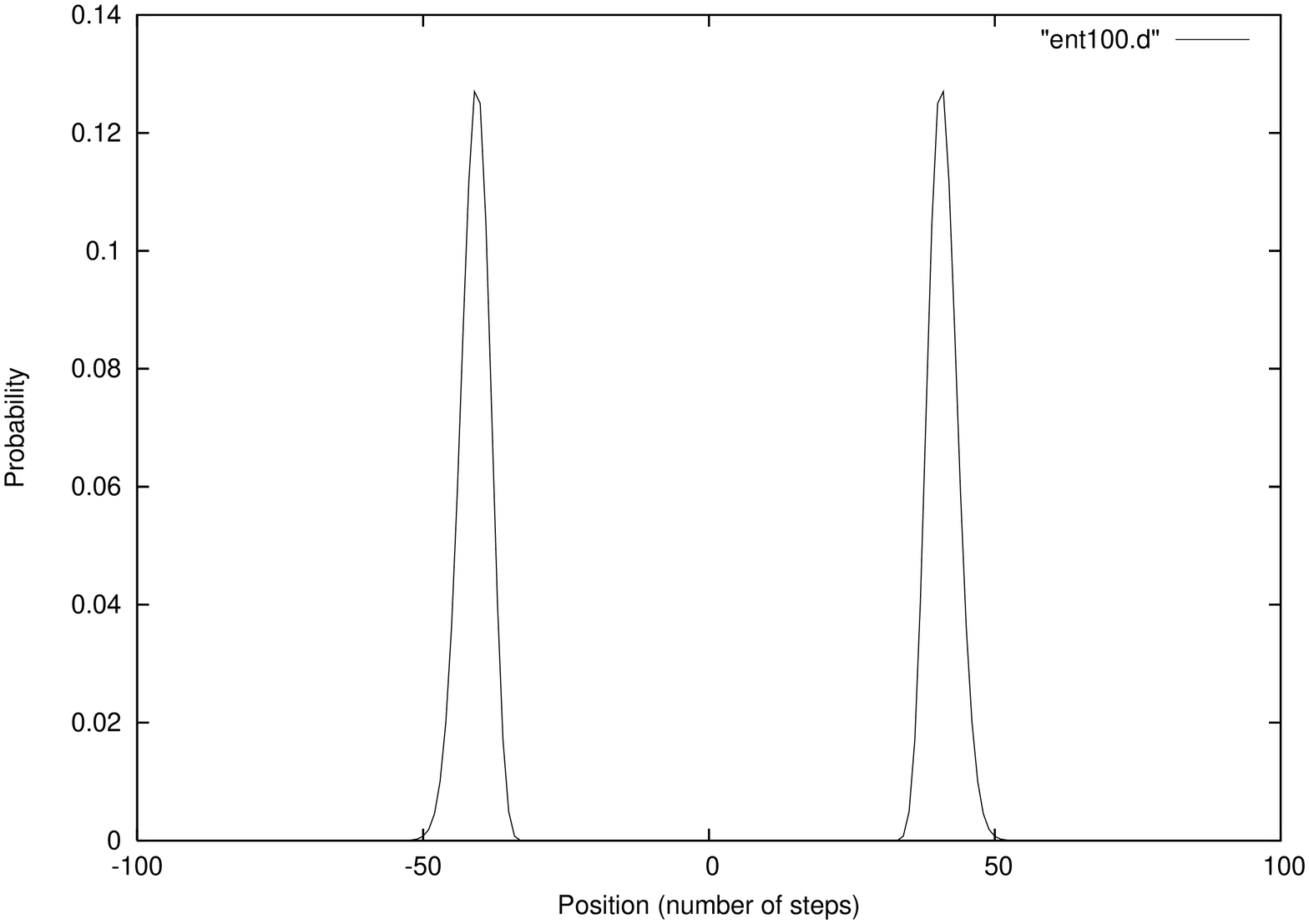, height= 9.4cm, angle=-90}
\caption{\label{entbec100}  Probability distribution of
the constrained quantum walk using entangled particle pair. The distribution is for 100 steps.}
\end{center}
\end{figure}

The combined unitary operator on the system can be written as, 

\be
U^{\prime}_{12} = U_{1}\otimes U_{2}.
\ee

If the initial state of the inter species two-particle entangled BEC can be written as,
\be \label{ent-pos0-bec} |\psi_{0}\rangle_{BEC} = \frac{1}{\sqrt 2}[|0, x_{0}\rangle_{1}|1, x_{0}\rangle_{2} + i |1, x_{0}\rangle_{1}|0, x_{0}\rangle_{2}]_{BEC}. \ee 
The state of the system after the first unitary operation can be written as,

\begin{eqnarray} 
\label{end-posx11-bec} 
U^{\prime}_{12} |\psi_{0}\rangle_{BEC} = \frac{1}{\sqrt 8}[(|0, x-1\rangle_{1}|1, x-1\rangle_{2} + i |1, x-1\rangle_{1} |0, x-1\rangle_{2})_{BEC} + \nonumber \\
(|0, x+1\rangle_{1}|1, x+1\rangle_{2} + i |1, x+1\rangle_{1} |0, x+1\rangle_{2})_{BEC} + 2(|0, x\rangle_{1}|1, x\rangle_{2} + i |1, x\rangle_{1} |0, x\rangle_{2})_{BEC}]. 
\end{eqnarray} 
From Eq. (~\ref{end-posx11-bec}) one can note that upon measurement after implementing the walk, the two entangled particles are always found together. In this special case the number of runs required to confirm the quantum behavior of the walker does not reduce as it does in general case discussed in Sec. \ref{entangled-particle}. 

The states of the inter species two-particle BEC, along with moving in superposition of position space without loosing the superposition in the coin space, a significant fraction (probability amplitude) of the state remains in the same place without any displacement. But with each unitary operator that fraction reduces and a quantum walk is implemented without loosing the superposition in the coin space. Fig. (\ref{entbec20}) and Fig. (\ref{entbec100}) are the simulation of the probability distribution of the entangled pair quantum walker with constrains as discussed for inter species entangled particle BEC. From the simulation one can note that the position of finding the state with maximum probability moves slowly as compared to the general case and the probability is highly concentrated around the maximum. Anyway this is one of the extreme example and for most of the physical systems the model constructed in Sec. \ref{entangled-particle} should suffice.  

%===================================
\section{Conclusion} \label{conclusion}
%===================================

In summary, we have constructed a quantum walk model for single particle and 
two two-state maximally entangled particles and have shown that after
every unitary shift operator $U^{\prime}$, the entanglement in the coin space of the particles before the beginning of the walk is retained. This preservation of
superposition state in the coin space eliminates the need for Hadamard operation used for entangling the coin space after every unitary operator in the most
widely studied version of discrete time quantum walk. The model
presented in this paper seems to be the simplest version for experimental realization of the quantum walk for both single particle and paired entangled particles and can be extended to mixed and multiple particle quantum walker. For entangled particle this construction also reduces the number of runs of quantum walk necessary to conclude the quantum behavior of the entangled particle walker and this can be of great advantage for the applications of quantum walk. Implementation of quantum walk on a pair of entangled particles without a coin toss on a complex physical system such as two species entangled particle BEC has also been treated as a special case.

%==========================
\bc
\bf Acknowledgments
\ec
%=============================
The author would like to thank Professor R. Simon for general comments on quantum information theory. Institute of Mathematical Sciences, Chennai and Harish-Chandra Research Institute, Allahabad, India for their hospitality during summer 2006.
Andris Ambainis, Ashwin Nayak, R. Laflamme and members of the Institute for quantum
computing for various fruitful discussions while writing this part of the work.

%=================================

\end{document}